\documentclass[conference,letterpaper,9pt]{IEEEtran}
\IEEEoverridecommandlockouts
\usepackage{cite}
\usepackage{amsmath,amssymb,amsfonts}
\usepackage{algorithmic}
\usepackage{graphicx}
\usepackage{textcomp}
\usepackage{xcolor}
\usepackage{tikz}

\newcommand{\circlednumber}[1]{%
    \tikz[baseline=(char.base)]{
        \node[shape=circle,draw,inner sep=1pt] (char) {#1};}}

\usepackage{fancyhdr}
\usepackage[hyphens]{url}

\usepackage{comment}
\usepackage{epstopdf}
\usepackage{amssymb}
\usepackage{tikz}
\usepackage{booktabs,caption,fixltx2e}
\usepackage[flushleft]{threeparttable}
\usepackage{dblfloatfix}

\usepackage{xcolor}
\definecolor{myBlue}{rgb}{0,0.3,0.5}

\usepackage{xspace}

\usepackage{lipsum}
\def\BibTeX{{\rm B\kern-.05em{\sc i\kern-.025em b}\kern-.08em
    T\kern-.1667em\lower.7ex\hbox{E}\kern-.125emX}}
\begin{document}

\title{Jack Unit: An Area- and Energy-Efficient Multiply-Accumulate (MAC) Unit Supporting Diverse Data Formats
}

\author{
Seock-Hwan Noh$^\ast$, Sungju Kim$^\dagger$, Seohyun Kim$^\ast$, Daehoon Kim$^\dagger$, Jaeha Kung\textsuperscript{\ddag\,\S} and Yeseong Kim$^\ast$\textsuperscript{\S} \\[1ex]
$^\ast$DGIST, Daegu, Republic of Korea \\
$^\dagger$Yonsei University, Seoul, Republic of Korea \\
$^\ddag$Korea University, Seoul, Republic of Korea \\[1ex]

nosh3332@dgist.ac.kr;
jhkung@korea.ac.kr;
yeseongkim@dgist.ac.kr
\thanks{\textsuperscript{\S}Jaeha Kung and Yeseong Kim are corresponding authors \{\textit{Email:} \underline{jhkung@korea.ac.kr} and \underline{yeseongkim@dgist.ac.kr}\}}
}

\maketitle


\begin{abstract}\small
In this work, we introduce an area- and energy-efficient multiply-accumulate (MAC) unit, named \textit{\underline{Jack Unit}}, that is a jack-of-all-trades, supporting various data formats such as integer (INT), floating point (FP), and microscaling data format (MX). 
It provides bit-level flexibility and enhances hardware efficiency by i) replacing the carry-save multiplier (CSM) in the FP multiplier with a precision-scalable CSM, ii) performing the adjustment of significands based on the exponent differences within the CSM, and iii) utilizing 2D sub-word parallelism. 
To assess effectiveness, we implemented the layout of the Jack unit and three baseline MAC units.
Additionally, we designed an AI accelerator equipped with our Jack units to compare with a state-of-the-art AI accelerator supporting various data formats. 
The proposed MAC unit occupies 1.17$\sim$2.01$\times$ smaller area and consumes 1.05$\sim$1.84$\times$ lower power compared to the baseline MAC units. 
On five AI benchmarks, the accelerator designed with our Jack units improves energy efficiency by 1.32$\sim$5.41$\times$ over the baseline across various data formats.
\end{abstract}


\section{Introduction}
Due to remarkable advancements in AI accuracy, many applications have begun to leverage AI algorithms for various tasks such as language translation, autonomous driving, and voice recognition. 
The success of AI in achieving high accuracy is driven by the processing of massive amounts of data.
For instance, OpenAI’s GPT-4 and Meta’s LLaMA handle multi-modal queries with high accuracy, utilizing models with parameters ranging from tens of billions to trillions~\cite{gpt4, llama}. 
However, managing such vast amounts of data using IEEE double-precision (FP64) or single-precision (FP32) formats, which are commonly employed in computer systems, results in excessive hardware costs, particularly in terms of memory footprint and energy consumption.
To alleviate this burden, algorithmic methods for reducing bit-width have been actively explored across diverse data formats, including integer (INT)~\cite{int8_training, understand_INT4, llm_8bit}, floating point (FP)~\cite{bfloat16, fp8_qualcomm, hybrid_fp8}, and microscaling (MX) formats~\cite{mx_format_nips, flexblock, mx_format_isca}.

\begin{figure}[t]
    \centering
    \includegraphics[scale=0.65]{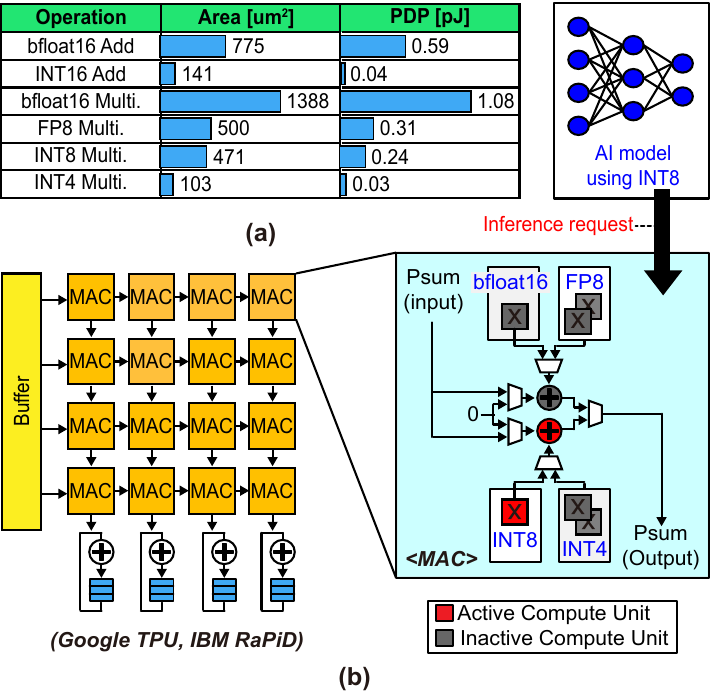} 
    \caption{
    (a) Hardware costs, i.e., area and power-delay product (PDP), for compute units across various data formats. These hardware costs are derived from Synopsys DesignWare IPs using a 65 nm CMOS technology. (b) Structure of a MAC unit in accelerators that deploy dedicated multipliers for each supported data format.
}    
    \vspace{-4mm}
    \label{fig:motivation_figure}
\end{figure}

AI-based applications exploit different precisions and data types based on the target devices and task objectives. 
Specifically, applications that require high accuracy typically adopt high precision to minimize accuracy loss, while those running on small devices, such as smartwatches and drones, may use low precision due to power and area constraints. 
Additionally, the training and inference processes of AI models have different requirements for data formats due to the varying dynamic ranges in data representations.
To meet these diverse requirements, many research efforts and commercial AI processors integrate multiple compute units supporting various data formats into a single chip (or chipset)~\cite{tpu_v4, rapid, nvidia_rtx, flexblock}.
However, this approach leads to low area efficiency. 
Fig.~\ref{fig:motivation_figure}-(a) illustrates the hardware costs, i.e., area and power-delay product (PDP), for several data formats that are widely used in quantized AI algorithms.
To support multiple data formats, placing as many compute units as supported precisions, as depicted in Fig.~\ref{fig:motivation_figure}-(b), increases area cost of multiply-accumulate (MAC) units.
Furthermore, the tendency of AI algorithms to typically use a single data format leaves some parts of the MAC units idle. 
As a result, accelerators that integrate multiple compute units have poor area efficiency. 
To address this issue, some research efforts employ multiple accelerators to accommodate the heterogeneity of AI models~\cite{nnp_t, brainwave}, but this increases development costs in both hardware and software stacks. 
This is because device drivers must be developed for each accelerator deployed in the heterogeneous system, and the system compiler needs to be updated to manage the multiple accelerator substrates~\cite{lynx}.

Considering these limitations, in this paper, we present an area- and energy-efficient MAC unit, named the \textbf{\textit{\underline{Jack unit}}}, which is a versatile solution for various data formats.
Specifically, the proposed MAC unit offers bit-level flexibility for INT, FP, and MX data formats (e.g., MXINT and MXFP) at a low hardware cost (\textit{\textbf{contribution of this work}}) by leveraging three key design strategies: i) replacing the carry-save multiplier (CSM) within the FP multiplier with a precision-scalable CSM, ii) performing the adjustment of significands within the CSM, and iii) utilizing 2D sub-word parallelism.
Replacing the multiplier with a precision-scalable CSM enables the multiplication of various data formats within a single integrated multiplier, eliminating the need for separate multipliers for each supported format (Sec.~\ref{sec:precision_scalable_CSM}). 
Performing the adjustment of significands in response to exponent differences within the CSM allows the accumulation of multiplication results using an INT adder tree instead of a FP adder tree (Sec.~\ref{sec:exp_adjust}).
Additionally, the 2D sub-word parallelism reduces the area and power consumption of the adder tree used in accumulation by sharing shifters among the INT adders and reducing the input bit-width of the adder tree (Sec.~\ref{sec:subword_parallel}).
Using these strategies, the Jack unit reduces area and power by 2.01$\times$ and 1.84$\times$, respectively, compared to a commercial MAC unit.

To further demonstrate the practical effectiveness of the Jack unit, we designed an AI accelerator incorporating the Jack units and compared it against a state-of-the-art commercial AI accelerator. 
We conducted architectural evaluations on five representative AI algorithms performing different tasks.
The accelerator designed with the Jack units achieves a 2.02$\times$ reduction in area and demonstrates 1.32$\sim$5.4$\times$ energy efficiency improvements across various data formats compared to the baseline accelerator.


\begin{figure*}[t]
    \centering
    \includegraphics[scale=0.63]{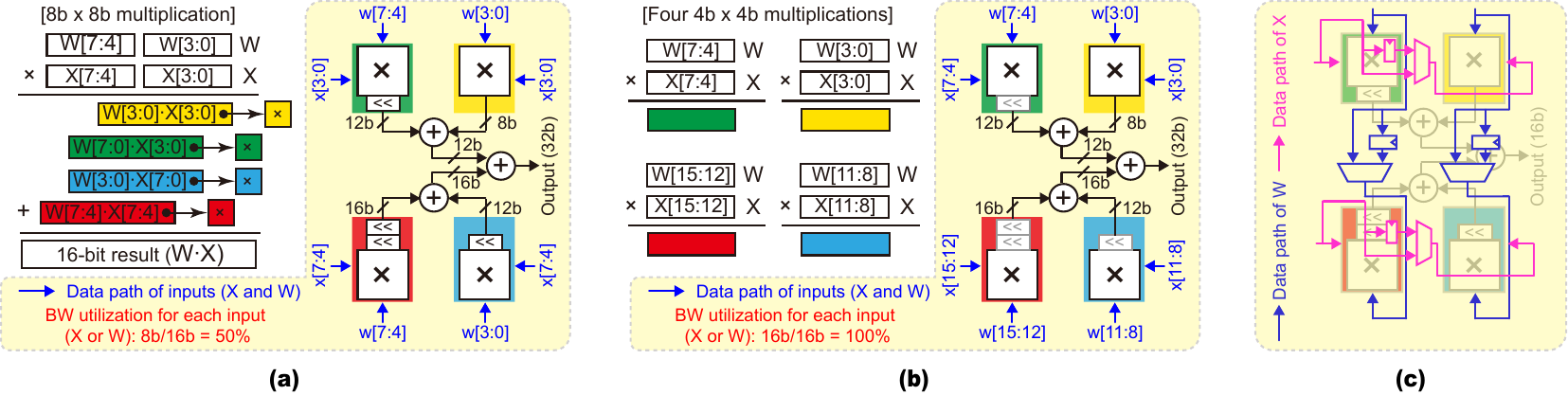} 
    \caption{Illustrations showing how the precision-scalable CSM operates on (a) 8-bit$\times$8-bit and (b) 4-bit$\times$4-bit multiplications. (c) Input datapaths of the reconstructed CSM in our design.}
    \label{fig:reconstructed_csm}
\end{figure*}

\begin{figure*}[!t]
    \centering
    \includegraphics[scale=0.65]{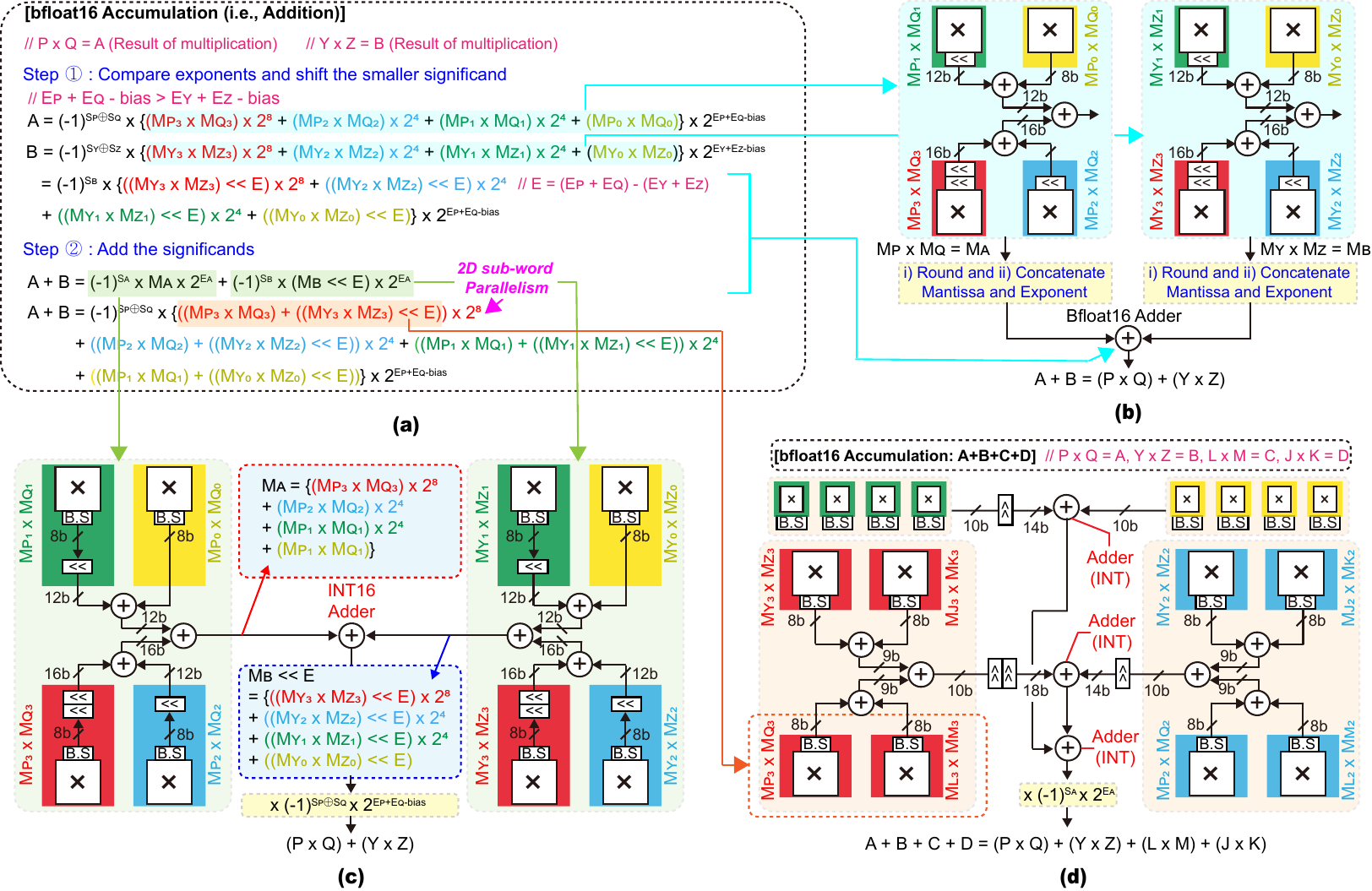}    
    \caption{(a) Computational flow of FP addition, (b) MAC unit that adds two multiplication results via a FP adder, (c) MAC unit that performs the adjustment of significands based on exponent differences within the CSM, (d) Reconstructed CSM used in our proposed Jack unit.}
    \vspace{-4mm}
    \label{fig:algorithm_fusion}
\end{figure*}

\section{Preliminaries}\label{sec:Preliminaries}

\subsection{Data Representation }\label{sec:data_representation}
Computer machines operate on two-state devices, meaning they use binary numbers, i.e., binary digits (bits), as their primary language. 
The computer system provides various data types to programmers by grouping binary numbers. 
Among these data types, one of the most commonly used numeric forms is the integer (INT) data type. 
In the INT numeral system, all digits of data represent whole numbers without fractional or decimal portions. 
An INT number with n bits can represent $2^{n}$ numbers; for example, an unsigned INT number can encode non-negative values from 0 through $2^{n}-1$, while a signed INT number has a value range of [$-2^{n-1}$, $+2^{n-1}-1$].
In contrast to the INT data type, the floating point (FP) data type, another commonly used data type, provides a fractional part by using a floating radix point. 
A FP number is represented as \vspace{-1mm}
\begin{equation}
    x_i = (-1)^{s_i} \times (1.m_i) \times 2^{e_i}, \vspace{-1.5mm}
\end{equation}
where $s_i$ is the sign, $m_i$ is the mantissa, and $e_i$ is the exponent of the number $x_i$.
Since this format can represent a wider dynamic range of numeric values than integers at the same bit-width, FP numbers are typically used in precision-sensitive applications that require a wide representable range of values. 
The FP data type can encode numbers with a value range of $\vert$$2^{B-1}$, $((2-2^{-M}))\times2^{B}$$\vert$, where $B$ is a bias value (calculated as $2^{E-1}$; $E$ is the bit-length of exponent), and $M$ is the bit-width of mantissa. 

Recently, several research efforts have applied microscaling (MX) data formats in AI algorithms~\cite{flexblock, mx_format_nips}. 
The MX format serves as an alternative representation that effectively balances logic density with the capacity to represent a wide range of values.
In this format, multiple numbers are grouped into a block, and these numbers share a common exponent value.
When the MX format numbers within a block are expressed as \vspace{-1.5mm}
\begin{equation}\label{eq:mx_format}
    \vec{x} = [\hat{x_1}, \hat{x_2}, \ldots, \hat{x_N}]\cdot 2^{e_{x}} = \vec{\hat{x}}\cdot 2^{e_{x}}, \vspace{-1.5mm}
\end{equation}
the dot product between two blocks is computed \vspace{-1mm}
\begin{equation}\label{eq:mx_dot_product}
    \vec{x}\cdot\vec{y} = (\vec{\hat{x}}\cdot\vec{\hat{y}})\cdot2^{(e_{x}+e_{y})}. \vspace{-1mm}
\end{equation} 
In Eq.~(\ref{eq:mx_dot_product}), the elements of $\vec{x}$ and $\vec{y}$ can be either INT or FP numbers, which are referred to as the MXINT format and the MXFP format, respectively.
By utilizing the MX format, even if the elements of $\vec{x}$ and $\vec{y}$ are quantized to low bit-width, the shared exponents (i.e., $2^{e_{x}}$ and $2^{e_{y}}$) enable the representation of wide dynamic ranges while allowing the dot product ($\vec{\hat{x}}\cdot\vec{\hat{y}}$) to be computed at a lower hardware cost.
Additionally, the memory footprint associated with storing exponents is reduced by sharing a common exponent among elements.
Given these significant advantages, several leading technology companies, such as Intel, Meta, Microsoft, and NVIDIA, are advancing the standardization of the MX format as a next-generation narrow precision data format for AI~\cite{ocp_block}.

\subsection{Related Work}\label{sec:related_work}
Research efforts aimed at reducing bit-width have led to the design of AI accelerators that support bit-level flexibility.
For example, several studies~\cite{unpu, loom} support variable precision configurations for the INT data type using bit-serial MAC units. 
In contrast, the works in~\cite{bit_fusion, art_mac} and~\cite{bit_blade} employ a spatial approach, merging results generated from sub-multipliers that construct MAC units.
Focusing on training of AI algorithms, MAC units capable of precision scaling in MX formats have also been studied~\cite{flexblock, bucket_bfp}. 
These MAC units offer low hardware costs and error-tolerant training results, due to simplified MAC operations and the ability to cover a wide spectrum of required precision. 
In hyperscale cloud infrastructures, AI accelerators typically support mixed precision for INT and FP, enabling cloud systems to handle both inference and training tasks on the same accelerator node~\cite{rapid, tpu_v4, nvidia_rtx}.
Recently, commercial AI hardware platforms, designed for mobile and portable devices, have incorporated bit-flexibility for INT or mixed precision by embedding custom-developed ASICs~\cite{samsung_flag, arm}. 
However, notably, there have been no efforts to integrate bit-level flexibility for INT, FP, and MX formats simultaneously (\textit{\textbf{unaddressed challenge}}). 
Most AI accelerators (e.g., Google TPU~\cite{tpu_v4}, IBM RaPiD~\cite{rapid}, and NVIDIA Tensor Core~\cite{nvidia_rtx}) offered by cloud service providers achieve bit scaling by deploying multiple compute units for each supported data format.

\section{Jack Unit}\label{sec:jac_unit}
Recent state-of-the-art advances in reducing bit-width have demonstrated successful training and inference accuracy across various data formats (e.g., bfloat16 \{s:1, e:8, m:7\}, FP8 \{s:1, e:4, m:3\}, INT8, INT4, MXINT8, MXINT4, and MXFP8 \{s:1, e:4, m:3\}\footnote{The data format is represented as \{s: sign bit, e: exponent bits, m: mantissa bits\} throughout this paper.}), along with improvements in both performance and energy efficiency~\cite{int8_training, understand_INT4, llm_8bit, bfloat16, fp8_qualcomm, hybrid_fp8, flexblock, mx_format_nips}. 
Inspired by these recent research efforts, we propose a novel MAC unit, named the Jack unit, which can support a wide range of data formats.

\begin{figure*}[t]
    \centering
    \includegraphics[scale=0.67]{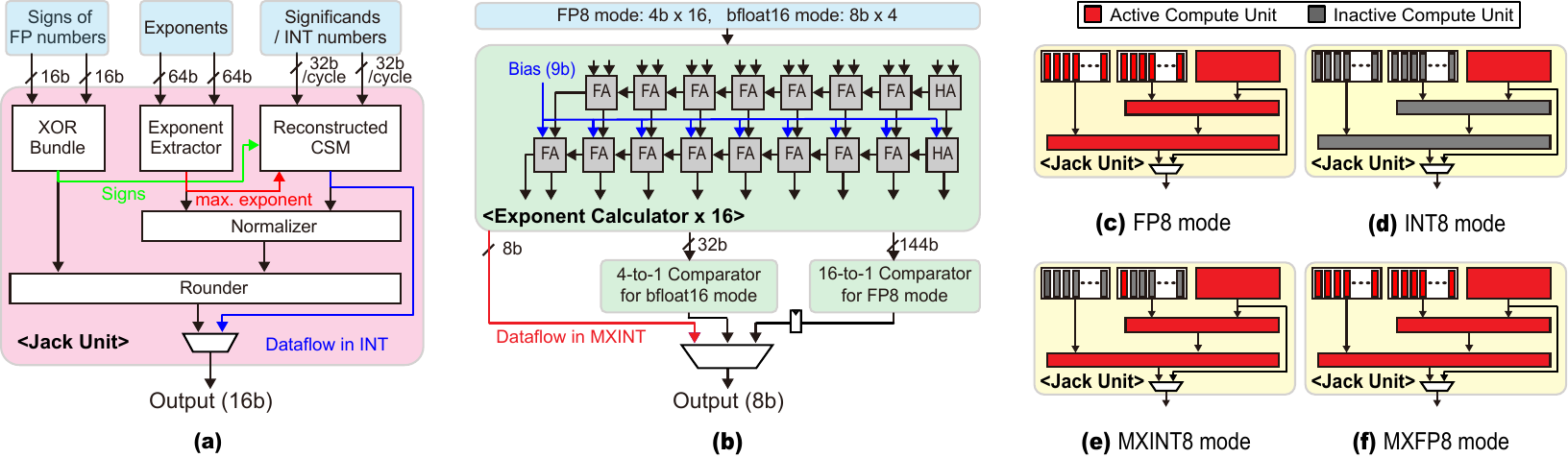} \vspace{-2mm}
    \caption{(a) Overall structure of the proposed Jack unit, (b) Structural diagram of an exponent extractor. (c-f) Activation of sub-modules in the Jack unit under different operation modes, i.e., FP8, INT8, MXINT8, and MXFP8 modes.}  
    \label{fig:structure_jack}\vspace{-4mm}
\end{figure*}

\vspace{-1mm}
\subsection{Reconstruction of Carry-save multiplier (CSM)}\label{sec:algorithm_fusion}

\vspace{-1mm}
\subsubsection{Bit-scalable CSM with Datapath Pipeline}\label{sec:precision_scalable_CSM}To reduce the hardware cost of the MAC unit, we first reconstruct the CSM of a FP multiplier. 
The multiplication result of two FP numbers is $x \times y = (-1)^{S_x \oplus S_y} \times (M_x \times M_y) \times 2^{E_x+E_y-Bias}$, where $M_x$ and $M_y$ are significands of $x$ and $y$ (i.e., $1.m_x$ and $1.m_y$), respectively.
The CSM is a sub-module of the FP multiplier, performing the multiplication of significands.
When synthesizing the bfloat16 and FP8 multipliers, which were introduced in the previous work~\cite{fp_multiplier}, the CSMs occupy the largest portion of area and power consumption among the sub-modules of the multipliers.
Specifically, under maximum operating frequencies, the CSMs account for 73.3\% of the area and 71.1\% of the power in the bfloat16 multiplier, and 53.8\% of the area and 47.3\% of the power in the FP8 multiplier. 
Thus, allocating dedicated multipliers for each supported precision level in a MAC unit, as depicted in Fig.~\ref{fig:motivation_figure}-(b), can significantly increase hardware cost due to the integration of such expensive CSMs for each supported data format.
To alleviate this inefficiency, we replace the CSM of the bfloat16 multiplier with a cheaper, precision-scalable multiplier capable of supporting signed/unsigned multiplications of 4-bit and 8-bit combinations, similar to the approach in~\cite{bit_fusion}. 
Fig.~\ref{fig:reconstructed_csm}-(a) and (b) show the structure of the replaced CSM with two examples illustrating how the precision-scalable CSM operates in 8-bit$\times$8-bit and 4-bit$\times$4-bit modes. 
As shown, the replaced CSM consists of four 4-bit$\times$4-bit sub-multipliers and supports 4-bit and 8-bit configurations by dynamically fusing the results of the sub-multipliers. 
However, this structure presents a design challenge due to varying bandwidth (BW) utilization depending on the supported precision level\footnote{High bandwidth (BW) can increase the wire area, leading to an increase in overall area. 
We performed P\&R with 90\% core utilization on two 8$\times$8 MAC arrays.
The arrays are designed with non-pipelined (Fig.~\ref{fig:reconstructed_csm}-(a-b)) and pipelined JACK units, respectively (Fig.~\ref{fig:reconstructed_csm}-(c)). 
We observed that the non-pipelined MAC array incurred approximately a 42\% layout overhead.}. 
For instance, the BW of inputs is fully utilized in the 4-bit precision mode (Fig.~\ref{fig:reconstructed_csm}-(b)), but only 50\% of the BW is utilized in the 8-bit precision mode (Fig.~\ref{fig:reconstructed_csm}-(a)).
To mitigate this problem, we modify the datapaths of the inputs into bypassable wired links, as depicted in Fig.~\ref{fig:reconstructed_csm}-(c). 
The bypassable link consists of 8 signal wires for each input (i.e., X and W) and guarantees 100\% BW utilization regardless of the supported precision level by delivering wider inputs ($>$ 8-bit) over multiple cycles through the pipelined path.

\subsubsection{Adjusting Significands in CSM}\label{sec:exp_adjust}
Fig.~\ref{fig:algorithm_fusion}-(a) presents the computational flow of FP addition, which adds two multiplication results. 
When adding FP numbers with different exponents, it is necessary to align the exponent of the smaller number with that of the larger. 
To achieve this, FP addition involves two major steps: i) comparing the exponents and shifting the significand of the smaller value by the exponent difference, and ii) adding the aligned significands. 
These steps are complex processes that do not exist in INT addition. 
Thus, an FP adder incurs higher hardware costs than an INT adder at the same bit-width (Fig.~\ref{fig:motivation_figure}-(a)).
Although the CSM has been replaced with a precision-scalable CSM, as discussed in the previous subsection, the MAC unit is still costly due to the inclusion of a FP adder (Fig.~\ref{fig:algorithm_fusion}-(b)).
To enable MAC operations at a much lower hardware cost, we explored an approach that allows the addition to be performed using an INT adder\footnote{
The Jack unit uses an INT adder for accumulation in FP modes without intermediate normalization or rounding, yet results in minimal error compared to standard FP MAC operations.
The absence of rounding reduces accumulation error, thereby maintaining numerical accuracy comparable to that of FP MAC computations.
To verify this, we implemented the Jack unit in C language and compared its results on the 2nd layer of ConvNeXt-T~\cite{convnext} with GPU results, which resulted in numerical errors within 0.2\%.}.
The core concept behind this method is to calculate the exponent difference in advance and shift the significand of the smaller value within the CSM.
Fig.~\ref{fig:algorithm_fusion}-(c) shows the block diagram of the MAC unit that supports this.
The outputs from the sub-multipliers are immediately shifted by barrel shifters (labeled as B.S. in the figure) before being fed into the intra-CSM adder tree. 
This approach enables both addition and multiplication to be performed on INT-only computation units.

\subsubsection{2D Sub-word Parallelism}\label{sec:subword_parallel}
To further reduce hardware costs, we rearrange the sub-multipliers to enable 2D sub-word parallelism on the operands.
In the precision-scalable CSM, left-shift operations are sequentially executed after the multiplications are performed by the sub-multipliers.
It is noteworthy that the outputs of sub-multipliers in the same positions across different CSMs share identical left-shift parameters.
For instance, as illustrated in Fig.~\ref{fig:algorithm_fusion}-(c), the outputs of sub-multipliers in the red boxes are shifted by the same value.
(Specifically, the sub-multipliers within the red boxes perform the multiplications of sub-words A and B (i.e., $M_{P_{3}} \times M_{Q_{3}}$ and $M_{Y_{3}} \times M_{Z_{3}}$) and are left-shifted by the same shift parameter, i.e., $2^8$, through barrel shifters.)
Therefore, we group the sub-multipliers that have the same shift parameters, allowing these groups to share left-shift operators.
Fig.~\ref{fig:algorithm_fusion}-(d) depicts the final structure of the reconstructed CSM utilized by the Jack unit.
This reconstructed CSM clusters sub-multipliers from four precision-scalable CSMs. 
With this approach, the reconstructed CSM reduces the number of shifters by 75\% compared to the four CSMs that do not use 2D sub-word parallelism.

\vspace{-1mm}
\subsection{Overall Structure of Jack Unit}\label{sec:structure_jack}\vspace{-1mm}
Figure~\ref{fig:structure_jack}-(a) shows the overall structure of our proposed Jack unit. 
The Jack unit consists of a reconstructed CSM, a XOR bundle, an exponent extractor, a normalizer, and a rounder. 
The reconstructed CSM serves as a building block that groups four bit-scalable CSMs. 
As described in Sec.~\ref{sec:precision_scalable_CSM}, a precision-scalable CSM performs a single 8-bit$\times$8-bit multiplication and four 4-bit$\times$4-bit multiplications when the significands of the operands have 8-bit (e.g., bfloat16) and 4-bit (e.g., FP8) lengths, respectively.
Thus, the Jack unit produces four multiplication results when the significands have a 8-bit length and sixteen multiplication results when they have a 4-bit length.
To match the number of arithmetic results produced by the reconstructed CSM, the XOR bundle and exponent extractor include sixteen XOR gates and exponent calculators in their internal structure. 
The XOR bundle calculates the signs of the data formats with a floating radix point (e.g., FP and MXFP), while the exponent extractor determines the maximum exponent value among the results of these formats using exponent calculators and comparators (Fig.~\ref{fig:structure_jack}-(b)). 
The results from the XOR bundle and exponent extractor are transferred to the reconstructed CSM and are utilized in the process of calculating the significands.
After completing the MAC operations in the reconstructed CSM, the maximum exponent value and the outputs of the reconstructed CSM are delivered to the normalizer. 
The normalizer detects the leading one value and normalizes the mantissas by incrementing the exponents if overflows occur. 
Finally, the rounder truncates the bit-width of the resultant data to a 16-bit length.
Note that the Jack unit produces a single 16-bit output, i.e., FP16 or INT16, to avoid accumulation errors, in line with the design presented in~\cite{rapid}.

\begin{figure}[t]
    \centering
    \includegraphics[scale=0.68]{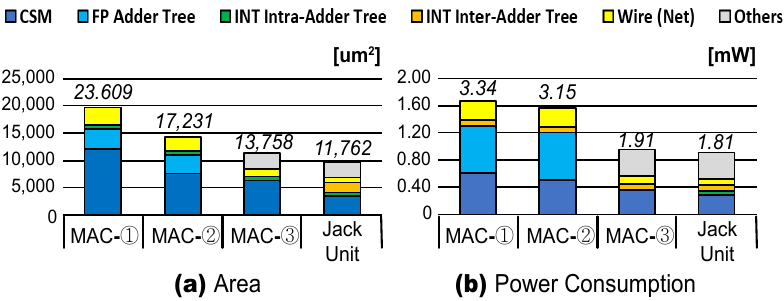} 
    \caption{(a) Area and (b) power breakdown of MAC units.}    
    \vspace{-2mm}
    \label{fig:mac_comparison}    \vspace{-2mm}
\end{figure}

\subsection{Operating Mode}\label{sec:operating_mode}\vspace{-1mm}
The proposed Jack unit supports flexible bit configurations for various data types, such as FP, INT, and MX data formats. 
This flexibility is achieved through the selective power gating scheme that deactivates the sub-modules within the Jack unit when they are not in use. 
Fig.~\ref{fig:structure_jack}-(c-f) illustrate how the sub-modules are activated in several supported data format modes, i.e., FP8, INT8, MXINT8, and MXFP8 modes.
In FP8 mode, all sub-modules of the Jack unit are activated, as shown in Fig.~\ref{fig:structure_jack}-(c).
In contrast, only the reconfigured CSM is activated in INT8 mode (Fig.~\ref{fig:structure_jack}-(d)). 
In this case, the result of the reconfigured CSM is directly routed to the output of the Jack unit via a dedicated datapath, as illustrated in Fig.~\ref{fig:structure_jack}-(a).
In MXINT8 mode, a single exponent calculator is activated, since the elements in the block tensors share a common exponent value.
In MXFP8 mode, similar to FP8 mode, all sub-modules are activated; however, the exponent calculator receives a bias that incorporates the shared exponents of the blocks, i.e., $2^{e_x}$ and $2^{e_y}$ in Eq.~\ref{eq:mx_dot_product}. 
(Specifically, in MXFP mode, the exponent calculator takes $bias$ + $e_x$ + $e_y$ as its bias value.)
This allows the Jack unit to perform computations on the shared exponents of the blocks within the unit itself.
Note that in the MX format mode, the granularity of the block size can be adjusted by activating the exponent calculators across multiple Jack units. 
For instance, in MXINT16 mode, sixteen elements can share a common exponent value by activating only one exponent calculator across four Jack units.

\section{Evaluation}\label{sec:evaluation}
\subsection{Delay, Area, and Power Analysis for the Jack Unit}\label{sec:jack_comparison}
\noindent\textbf{Baselines and Methodology: }To quantitatively evaluate the implications of our approaches, we implemented the Jack unit and three baselines at the register transfer level (RTL). 
One of the baselines is a MAC unit (MAC-\scalebox{0.9}{\circlednumber{1}}) that supports various data formats by incorporating dedicated multipliers for each supported data format, similar to commercial AI accelerators~\cite{tpu_v4, rapid, nvidia_rtx}.
The other two baselines, i.e., MAC-\scalebox{0.9}{\circlednumber{2}} and MAC-\scalebox{0.9}{\circlednumber{3}}, are the MAC units depicted in Fig.~\ref{fig:algorithm_fusion}-(b) and (c). 
The MAC-\scalebox{0.9}{\circlednumber{2}} is a MAC unit in which the CSM is replaced with a precision-scalable CSM, while MAC-\scalebox{0.9}{\circlednumber{3}} not only replaces the CSM with a precision-scalable CSM, but also performs the adjustment of significands within the CSM. 
Driven by recent algorithmic advancements in quantization~\cite{int8_training, understand_INT4, llm_8bit, bfloat16, fp8_qualcomm, hybrid_fp8}, the baselines support data formats such as bfloat16, FP8, INT8, and INT4 with the same throughput as the proposed Jack unit.
All designs were synthesized using the typical PVT corner (1.1 V, 25 $^\circ$C) of a 65 nm CMOS library with Synopsys Design Compiler~\cite{synopsys_dc}.
For the delay, we report the minimum achievable clock periods that do not cause setup time violations.
We present the area data obtained from Synopsys IC Compiler~\cite{synopsys_icc} and the power data estimated using Synopsys PrimeTime PX~\cite{pt_synopsys}, all based on placement and routing (P\&R) results under the same timing constraint of 286 MHz for a fair comparison.

\begin{table}[t]
\centering
\caption{Design configurations of the accelerator designed with our Jack units and a baseline accelerator (i.e., RaPiD-like accelerator~\cite{rapid}).}

\label{tab:design_configuration}
\scalebox{0.83}{%
\begin{tabular}{|c|c|c|}
\hline
\textbf{AI Accelerator}                                                                           & \textbf{Acc. w/ Jack units}                                                                           & \textbf{RaPiD-like Acc.}                                                                 \\ \hline\hline
\textbf{Technology Node}                                                              & Commercial 65 nm                                                                                                           & Commercial 65 nm                                                                                              \\ \hline
\textbf{Operating Frequency}                                                            & 400 MHz                                                                                                          & 400 MHz                                                                                             \\ \hline
\textbf{Array Size}                                                                 & 32$\times$32                                                                                                        & 128$\times$128                                                                                         \\ \hline
\textbf{Number of Multipliers}                                                          & \begin{tabular}[c]{@{}c@{}}128$\times$128  \\ (\texttt{bfloat16} / \texttt{INT8} / \texttt{MXINT8})\\ 512$\times$512  \\ (\texttt{FP8} / \texttt{INT4} / \texttt{MXFP8} / \texttt{MXINT4})\end{tabular} & \begin{tabular}[c]{@{}c@{}}128$\times$128 \\ (\texttt{bfloat16} / \texttt{INT8})\\ 512$\times$512 \\ (\texttt{FP8} / \texttt{INT4})\end{tabular} \\ \hline
\textbf{\begin{tabular}[c]{@{}c@{}}On-chip Buffer \\ (I / W / O) {[}KB{]}\end{tabular}} & 512 / 512 / 256                                                                                                  & 512 / 512 / 256                                                                                     \\ \hline
\end{tabular}
}
\vspace{+1mm}
\end{table}

\begin{table}[t]
\centering
\caption{AI benchmarks and datasets used for evaluation.}
\label{tab:benchmark_table}
\scalebox{0.89}{%
\begin{tabular}{|c|c|c|}
\hline
\textbf{Model}          & \textbf{Application}     & \textbf{Dataset} \\ \hline\hline
\textbf{ConvNeXt-T (CNN)~\cite{convnext}} & Image Classification     & ImageNet~\cite{imagenet}         \\ \hline
\textbf{BERT (NLP)~\cite{bert}}     & Language Modeling        & WMT14~\cite{wmt14}            \\ \hline
\textbf{GPT2-Small (LLM)~\cite{gpt2}}     & Language Modeling  &  Wikitext-2~\cite{wikitext2}           \\ \hline
\textbf{NeRF~\cite{nerf}}           & View Synthesis for AR/VR & Synthetic NeRF~\cite{synthetic_nerf}   \\ \hline
\textbf{QuickSRNet~\cite{QuickSRNet}}     & Image Super-Resuolution  & DIV2K~\cite{div2k}            \\ \hline
\end{tabular}
}
\vspace{+1mm}
\end{table}

\begin{figure}[!t]
    \centering
    \includegraphics[scale=0.85]{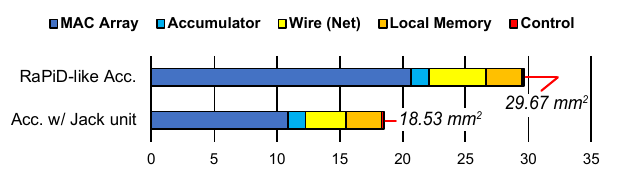} 
    \caption{Area breakdowns of the accelerator designed with Jack units and the baseline accelerator (i.e., RaPiD-like accelerator~\cite{rapid}).} 
    \label{fig:acc_area}\vspace{-2mm}
\end{figure}

\begin{figure*}[!t]
    \centering
    \includegraphics[scale=0.83]{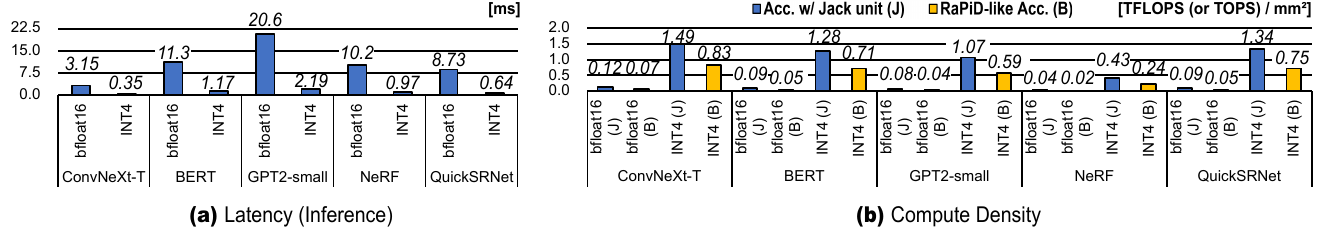} 
    \caption{(a) Inference latency of the JACK unit-based accelerator in bfloat16 and INT4 modes. (b) Comparison of compute density (i.e., area efficiency) between the Jack unit-based AI accelerator (J) and a baseline accelerator (B), under two precision modes: bfloat16 and INT4. 
    }
    \label{fig:latency_compute_density}\vspace{-2mm}
\end{figure*}

\noindent\textbf{Delay Analysis: }The MAC-{\circlednumber{1}} can operate without timing violations at clock frequencies up to 286 MHz ($T_{clk}$ = 3.5 ns).
The MAC-{\circlednumber{2}} has a longer delay of 3.6 ns than the MAC-{\circlednumber{1}}, due to an increase in the critical path caused by replacing the CSM with a bit-scalable CSM.
However, the MAC-{\circlednumber{3}} achieves an improved delay of 3.4 ns by adjusting the significands directly within the CSM, eliminating the need for accumulation through the FP adder tree and utilizing the cheaper INT adder tree instead.
The Jack unit further reduces the delay to 3.3 ns by employing 2D sub-word parallelism, which decreases the input bit-width of the INT adder tree.
This reduction in bit-width shortens the carry propagation delay within the accumulation adder, resulting in a shorter critical path.

\noindent\textbf{Area and Power Analysis: }Fig.~\ref{fig:mac_comparison} shows the area and power consumption of the baseline MAC units and the Jack unit. 
The MAC-{\circlednumber{1}}, which employs dedicated multipliers for each supported data format, occupies approximately 11,084um$^2$ of area and consumes 1.67mW of power. 
The MAC-{\circlednumber{2}} achieves a 1.37$\times$ reduction in area and a 1.06$\times$ in power consumption compared to the MAC-{\circlednumber{1}} by utilizing a CSM supporting bit-flexibility instead of dedicated multipliers. 
The hardware costs are further reduced by performing the adjustment of significands required for FP addition within the precision-scalable CSM.
The MAC-{\circlednumber{3}} incorporates this enhancement, reducing area and power by 20.15\% and 39.23\%, respectively, compared to the MAC-{\circlednumber{2}}.
Our proposed Jack unit demonstrates the smallest area and lowest power consumption, achieving a 1.17$\sim$2.01$\times$ reduction in area and a 1.05$\sim$1.84$\times$ reduction in power compared to the baseline units.

\begin{figure}[!t]
    \centering
    \includegraphics[scale=0.84]{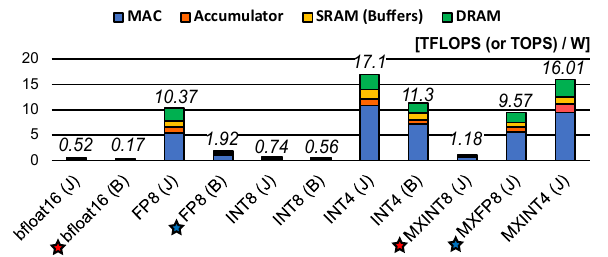} 
    \vspace{+1mm}
    \caption{Average energy efficiency of the JACK unit-based accelerator (J) and the baseline accelerator (B) across target AI benchmarks for each supported precision mode.} 
    \label{fig:energy_efficiency}\vspace{-4mm}
\end{figure}

\subsection{Architectural Evaluation}
\noindent\textbf{Baselines and Methodology: }We evaluated the practical effectiveness of Jack units by implementing a 32$\times$32 2-dimensional systolic array at the RTL.
For comparison, we implemented a recent AI accelerator, i.e., IBM RaPiD~\cite{rapid}, which features 128$\times$128 MAC units and provides the same throughput as the accelerator with Jack units across all its supported data format modes.
The Jack unit and the MAC unit employed in the both accelerators are pipelined with 2-stage registers to achieve the target clock frequency of 400 MHz.
To perform an architectural evaluation, we performed P\&R using the same EDA tools and the CMOS library as described in Sec.~\ref{sec:jack_comparison}. 
Additionally, power consumption for each supported data format was obtained by adjusting the switching activity of control signals in the power analysis tool, i.e., Synopsys PrimeTime PX. 
To conduct a system-level analysis, we considered both on-chip and off-chip memory accesses when estimating energy consumption. 
On-chip memory blocks were modeled using CACTI-6.0~\cite{cacti6}, while a dual-channel HBM was used for the off-chip power and timing specifications~\cite{hbm2}.
To estimate the clock cycles for both two architectures, we utilized an open-source cycle-level simulator, SCALE-sim~\cite{scale-sim}. 
We modified the simulator to reflect the on-chip and off-chip memory configurations and dataflows of the two architectures. 
Table~\ref{tab:design_configuration} summarizes the key design specifications of the two architectures.
Table~\ref{tab:benchmark_table} presents the list of evaluated AI benchmarks along with their corresponding datasets.


\noindent\textbf{Area Comparison: }Fig.~\ref{fig:acc_area} presents the area breakdowns of the accelerator designed with Jack units and RaPiD-like accelerator. 
The MAC array and wire areas of the Jack unit-based accelerator are 1.93$\times$ and 1.42$\times$ smaller than the corresponding areas of the baseline accelerator, respectively.
These reductions are attributed to the area-efficient design of the Jack unit and the pipelined datapath.
As a result, the Jack unit-based accelerator achieves an overall area that is 1.60$\times$ smaller than the baseline accelerator.

\noindent\textbf{Latency and Compute Density Analysis: }Fig.~\ref{fig:latency_compute_density}-(a) shows the inference runtimes for target benchmarks under two precision modes: bfloat16 and INT4. In INT4 mode, the Jack unit includes 16$\times$ more multipliers than in bfloat16 mode (Table~\ref{tab:design_configuration}), achieving inference speeds 9.06$\sim$13.08$\times$ faster. 
However, the Jack unit-based accelerator incurs 69\% higher on-chip buffer access latency due to the pipelined datapath, resulting in an overall 6.65\% longer inference latency compared to the baseline accelerator.
Despite the longer latency, the proposed JACK unit-based accelerator exhibits higher performance per unit area compared to the baseline accelerator.
Fig.~\ref{fig:latency_compute_density}-(b) presents the computational efficiency per area of both accelerators, showing that the MAC units composed of JACK units achieve, on average, 1.80$\times$ higher compute intensity.
This improvement stems from the area-efficient design strategies of the JACK unit, as discussed in Section~\ref{sec:algorithm_fusion}, along with the datapath pipeline.


\noindent\textbf{Energy Efficiency Analysis: }Fig.~\ref{fig:energy_efficiency} presents the energy efficiency of the two accelerators across various supported precision modes. 
Owing to the energy-efficient design of the proposed Jack unit, the Jack unit-based accelerator achieves 1.32$\sim$ 5.41$\times$ higher energy efficiency than the baseline accelerator under the identical bit-width configurations in both INT and FP modes.
Furthermore, unlike the baseline accelerator, the Jack unit-based accelerator supports MX formats\footnote{In our evaluation, we used a block size of 32, which is currently under standardization in the MX format~\cite{ocp_block}.}.
As indicated by the data marked with red stars in the figure, the Jack unit-based accelerator achieves 7.13$\times$ higher energy efficiency in MXINT8 \{s:1, e\_shared:8, m:7\} mode, compared to the baseline accelerator operating in bfloat16 \{s:1, e:8, m:7\} mode.
This improvement results from the use of low-precision multiplication enabled by the MX format, as discussed in Section~\ref{sec:Preliminaries}.
The data marked with blue stars in the figure show the energy efficiency of the Jack unit-based accelerator operating in MXFP8 \{s:1, e\_shared:8, e\_local:4, m:7\} mode and the baseline accelerator in FP8 \{s:1, e:4, m:3\} mode.
Although the Jack unit operates at a wider bit width, it achieves 4.98$\times$ higher energy efficiency due to its energy-efficient design.

\vspace{-2mm}
\section{Conclusion}\label{sec:conclusion}
This work introduces the Jack unit, an area and energy-efficient MAC unit supporting bit-flexible data formats through three design strategies: i) a precision-scalable multiplier, ii) significand adjustment within the CSM, and iii) 2D sub-word parallelism.
With these approaches, the proposed MAC unit achieves 2.01$\times$ reduction in area and a 1.84$\times$ decrease in power consumption compared to a MAC unit deployed in commercial AI accelerators. 
We expect that the Jack unit will be widely adopted in AI accelerators due to its ability to meet diverse design requirements and applications in a cost-effective manner.

\vspace{-1mm}
\section{Acknowledgment
}

This work was supported by the Institute of Information \& Communications Technology Planning \& Evaluation (IITP) grant funded by the Korean government (MSIT) under Grant No. 2022-0-00991, and by the National Research Foundation of Korea (NRF) grant funded by the Ministry of Science and ICT under Grant No. NRF-2023R1A2C2006290. It was also supported by the Technology Innovation Program (RS-2024-00445759, ``Development of Navigation Technology Utilizing Visual Information Based on Vision-Language Models for Understanding Dynamic Environments in Non-Learned Spaces") funded by the Ministry of Trade, Industry \& Energy (MOTIE, Korea). The EDA tool used in this work was supported by the IC Design Education Center (IDEC) in South Korea.

\bibliographystyle{plain}
\bibliography{ref}

\end{document}